**Grooming of Dynamic Traffic in WDM Star and Tree Networks Using Genetic Algorithm**

Kun-hong Liu[1,2], Yong Xu[3], De-shuang Huang[1], Min Cheng[4]

[1] Intelligent Computing Lab, Hefei Institute of Intelligent Machines, Chinese Academy of Sciences,
P.O. Box 1130, Hefei, Anhui, 230031, China.

[2] Department of Automation, University of Science and Technology of China,
Hefei, Anhui, 230026, China.

[3] School of Computer Science, The University of Birmingham
Edgbaston, Birmingham, B15 2TT, United Kingdom

[4] Department of Physics and Electronic Information Engineering, Minjiang University,
Fuzhou, 350108, China

Emails: lkhqz@163.com (K.H. Liu); y.xu@cs.bham.ac.uk (Y. Xu); dshuang@iim.ac.cn (D.S. Huang); cm94@163.com (M. Cheng)

**Abstract**

The advances in WDM technology lead to the great interest in traffic grooming problems. As traffic often changes from time to time, the problem of grooming dynamic traffic is of great practical value. In this paper, we discuss dynamic grooming of traffic in star and tree networks. A genetic algorithm (GA) based approach is proposed to support arbitrary dynamic traffic patterns, which minimizes the number of ADM's and wavelengths. To evaluate the algorithm, tighter bounds are derived. Computer simulation results show that our algorithm is efficient in reducing both the numbers of ADM's and wavelengths in tree and star networks.

## 1. Introduction

The advent of Wavelength Division Multiplexing (WDM) technique satisfies the ever-increasing bandwidth requirements of network users by carrying traffic of the order of Tbps. In WDM networks, it has become apparent that the cost of network component, especially add/drop multiplexers (ADM's) is one of most dominant factors. So it is important to reduce the number of ADM's by to intelligently assigning traffic flows to wavelengths, which is referred to as the traffic grooming (TG) problems.

Most of the previous work considered TG problems in ring or mesh topologies [1-10]. The early work mainly focused on rings [1-6] due to its widespread use in today's infrastructural networks. With the evolution of optical networks from ring to irregular topologies, the grooming of traffic in mesh networks is becoming a prominent research area [7-10] as mesh topology is more general and practical, especially in IP/WDM networks where packages are usually routed in such networks. However, other irregular topologies are also of great importance in their own sense. For example, star networks are widely deployed in the interconnection of LANs or MANs with a wide area backbone, whereas most cable TV networks and PONs are based on tree topologies. So it is also of great value to discuss TG problems in these topologies. But to the best of our knowledge, there are very few papers discussing about TG problems in stars and trees [3, 11-14].

The TG problem in star network is NP-complete even when the traffic demands remained unchanged during the grooming process, as was proved in [12]. It would be much harder when the grooming of arbitrary all-to-all dynamic traffic are considered in star topology, let alone in tree topology. So most researches mainly considered the TG problem based on the static traffic assumption [3, 11-14]. But this is not the best description of the traffic demands in most cases. Actually, the traffic pattern changes from time to time, especially in today's proliferation of IP traffic patterns. So a better description of the traffic would be a set of dynamic traffic patterns. The grooming of such traffic patterns is known as the grooming of dynamic traffic, which was first addressed by Berry and Modiano [4].

For the grooming of dynamic traffic, there are mainly two grooming categories [5]: *strictly nonblocking* and *rearrangeably nonblocking grooming*. To groom traffic in strictly nonblocking way is to assign the same traffic demand in different traffic patterns to the same wavelength, which allows all traffic demands in a new traffic pattern to be established without being interrupted. On the contrary, rearrangeably nonblocking grooming means that each traffic demand can be assigned to the different existing wavelengths with the set of nodes each wavelength drops at kept unchanged for each new traffic pattern, and it may result in saving more ADM's compared to the grooming of strictly nonblocking since more ADM's can be shared by rearranging some calls in different traffic patterns. But some of the existing traffic demands may be interrupted when new traffic has to be established. We will mainly focus on the strictly nonblocking grooming in WDM star and tree networks in this paper; the grooming of traffic in a rearrangeably nonblocking way will be our future work.

The TG problem in trees was proved to be NP-complete even when every interior node possessed full wavelength conversion capability [13]. The authors in [13] first decomposed a tree topology into stars, then used the algorithm to tackle the TG problem in star networks. It is obvious that when a decomposition method is employed, the algorithm would inevitably be more complex and time-consuming. What is more, they designed a greedy heuristic to solve the TG problem, in which the results would depend on the order of traffic elements being routed. Under this condition, it is hard to get globe optimality for all the cases even though the solution in each star reaches its optimal. Therefore, we have proposed a GA method which can groom traffic in tree topology with only one step with the order of traffic demands being no influence on the results in [14]. In this paper, we will improve the algorithm in [14], which can only be applied to the grooming of static traffic in tree networks, to deal with the grooming of arbitrary dynamic traffic in tree and star networks. Different to the goal in Ref. [12], which is to minimize the total amount of electronic switching, our objective is to minimize the total number of ADM's with as few wavelength as possible being used on the network.

The rest of this paper is organized as follows. We give the problem definition, discuss the TG problem in trees and stars respectively, and address the tighter upper and lower bounds in tree networks with some discussions in Section 2. In Section 3, we describe our new heuristic algorithms. Section 4 gives the computer simulation results along with corresponding discussions. In the last Section, we conclude this paper.

## 2. Problem Definition

In this paper, we propose a novel genetic algorithm to deal with the strictly nonblocking grooming in star and tree networks. The networks considered in this paper consist of $n$ nodes numbered by 0, 1, …, $n$-1. Each node is equipped with a number of SONET add/drop multiplexers (ADM's) which can aggregate a number of low-rate traffic streams onto a high-rate circuit. The dynamic traffic demands are represented by a set of $n \times n$ traffic matrix $\mathbf{R}=\{\mathbf{R}^m\}$ ($m$=0, 1, ..., $M$-1), each of which represents the traffic demands at time $t=m$. Each element $r_{ij}^m$ of the traffic pattern $\mathbf{R}^m$ represents the traffic demand originating from node $i$ and terminating at node $j$ at time $t=m$. Only one traffic pattern in this set is activated each time and all the $M$ traffic patterns will be constructed one by one as traffic changes. The wavelengths and ADM's used on the network must be able to allow each of these traffic patterns to be established each time after appropriate grooming. In [6], we proved that by splitting traffic, the usage of ADM's and wavelength can be further reduced. But in this paper, we do not allow the split of traffic in star and tree networks so as to simplify the problem. Discussion on traffic split in such topologies will be our future work.

The following discussions are made under the assumption that the maximum traffic demands are not larger than the capacity of a wavelength. It is reasonable because when a traffic demand $r_{ij}$ is larger than the granularity ($g$ for short) of a wavelength, it can be split into $\lfloor r_{ij}/g \rfloor$ full wavelength subcomponents, each of which can be assigned to a separate wavelength and one subcomponent with the magnitude of $r_{ij} - \lfloor r_{ij}/g \rfloor \times g$ less than $g$, which we would take into consideration when grooming traffic.

*2.1 Traffic Grooming in Tree Networks*

Tree networks can be classified into to categories: regular tree and irregular tree (random tree networks). The topology of regular tree networks takes on symmetry to some degree, in which binary tree is a good example. On the contrary, irregular tree networks are those whose topology has no symmetry.

In tree networks, there are two kinds of nodes: internal nodes and leaf nodes. As the internal nodes need to support two important functions: one is wavelength routing and the other is multiplexing and demultiplexing, these nodes should be equipped as those in mesh networks. We use the node architecture 3 defined in [7] to equip all the internal nodes, and equip each leaf node with a wavelength add/drop multiplexer (WADM). In this way, each internal node and leaf node should be only equipped with an ADM for each wavelength if the traffic needs to be added/dropped at that node. In our topology structure, all the links are bidirectional, that is, two fibers are installed between a connected father-child node pairs to support optical signal in both directions. We define the direction from father node to child node as direction 1, and the other direction, from child node to father node, as direction 2. In this way, the two directions at the left side of the root node, which we define as node 0, are opposite to those at the right side. To make our algorithm work efficiently, the number of nodes on each side should not be different too much so as to assure that the traffic demands are balanced in either side. To achieve the balance, we define the root node according to whether the number of nodes at the right side of a node is equal or close to the number at the left side.

In bidirectional ring and mesh networks, the choice of the links should be taken into consideration in order to route traffic on the networks. But in the tree networks, there is only one route between each node pair. This is significantly different from bidirectional ring and mesh networks and therefore, we can safely ignore the routing problem. If a traffic demand $r_{ij}$ is going to be assign to a wavelength, we only need to take the following two steps:
1. find the route between node $i$ and $j$;
2. judge whether all the links among the route have spare capacity to accommodate this traffic.

Step 1 tries to set up a lightpath between the node pair. If the relationship between each pair of nodes on the tree can be clearly defined, then by tracing a node's father or child node, a route can be established. In detail, after defining node 0, the position of each node in the tree can be consequently defined, then the corresponding information about a node, such as its layer, parent and child nodes (if it has) can also be recorded. After analyzing the properties of a node pair, we can establish the route by setting up a minimum subtree which contains these two nodes. For all kinds of tree networks, the process of establishing a route is similar. So the way of constructing of a tree topology will not affect the performance of our algorithm, and the algorithm can be applied to all kind of tree networks with slight adjustment in the description of the tree topology. For simplification, we present our algorithm in binary tree networks.

At the second step, if one of the links on the route exceeds its maximum load when the wavelength tries to accommodate the corresponding traffic demand, the traffic can not be assigned to that wavelength. What is more, a lightpath between internal nodes may carry some bypassed traffics as well. So when a lightpath is to be set up between an internal node and a leaf node, we should check the ADM's capacity when adding or dropping traffic to the current wavelength at the internal node. If a traffic cannot be added or dropped at the internal node, the corresponding lightpath cannot be setup either. In this case, a new wavelength should be used to accommodate the traffic request. As for a leaf node, such checking is unnecessary. After the checking, if it is found that there is enough spare space in the links, a traffic can be assigned to a wavelength.

*2.2 Traffic Grooming in Star Networks*

In star networks, only the hub node, which we label as node 0, is capable of switching traffic optically, and all other nodes do not need to be equipped with such a function. So we regard the star networks as a special kind of regular tree with only one internal node.

There are three kinds of traffic demand in star networks: from a leaf node to the hub node, from the hub node to a leaf node and from a leaf node to another leaf node. The first and the second kinds of traffic demand can be sent or received directly because the hub node is connected to every leaf node through a physical link. But the third kind of traffic must be routed at the hub node because both the two leaf nodes are only connected indirectly through the hub node. With the routing capability of the hub node, the traffic can be transferred without routing algorithms. So it is unnecessary to discuss the establishment of a route when assigning traffics to a wavelength in star networks, and we may only take the second step of assigning traffic in tree networks to estimate whether the links have enough spare capacity to accommodate a traffic demand.

Based on the property of star networks, we can tackle the TG problems in star networks with slightly adapting the algorithm designed for tree networks.

*2.3 Bounds on Irregular network*

The grooming problems in WDM irregular networks can be formulated as a set of multi-objective integer nonlinear programming equations. The primary objective is to minimize the number of ADM's, and the secondary objective is to minimize the number of wavelengths. As bounds on the number of ADM's and wavelengths will not only help to find the reasonable solution of the programming equations, but also be of great importance to evaluate the results of grooming, we will derive lower and upper bounds in the following subsections.

*2.3.1 Lower Bounds on the Irregular Networks*

Assume that two nodes *i* and *j* are directly connected and form a father-child node pair, in which *i* is the father and *j* the child. We refer to $L_{ij}^{1m}$ as the link load in the links which carry traffics from node pair *i* to *j* for traffic pattern *m* in direction 1, and $L_{ij}^{2m}$ as the link load in the links carrying traffic from node *j* to *i* in direction 2. Define the maximum link load of all the traffic patterns in the two directions between node pair *i* and *j* as $L^1_{ij}=\max_{x=0,\ldots,M-1}(L_{ij}^{1x})$ and $L^2_{ij}=\max_{x=0,\ldots,M-1}(L_{ij}^{2x})$ respectively.

In [14], the minimum number of wavelengths, denoted by $W_{\min}$, was set to be equal to the maximum number of wavelengths needed to carry the traffic in a lightpath. However, it should be noticed that only *g* traffic demands can be added or dropped at a node at most on a wavelength. Therefore, sometimes, the traffic demands that originate from the same internal node can be carried by the links connected to that node using the same wavelength, but can not be added at the internal node because the total traffic demands is larger than an ADM's capability. In this case, even when the links connected to the internal node can carry all the traffics, it will not guarantee that all the traffics can be safely originated or terminated at that node if there were not enough ADM's. For a leaf node, there is not such a problem as each leaf node is only connected to its father node. When a link carries *g* traffic demands they can all be assigned to the same wavelength.

Based on the observations above, the capacity of both the lightpaths and the ADM's must be taken into account when deriving the lower bounds on the number of wavelengths and ADM's. Let $\sigma^m(s)$ denotes the total traffic for traffic pattern *m* dropped at an internal node *s*, and $\tau^m(s)$ denotes the total traffic added at this node, then by defining $\sigma(s)=\max_{x=0,\ldots,M-1}(\sigma^x(s))$, $\tau(s)=\max_{x=0,\ldots,M-1}(\tau^x(s))$, and the set of the internal nodes in the tree as $\lambda$, the minimum number of wavelength is given by

$$W_{\min} = \max_{i,j=0,\ldots,n-1; i \neq j, s \in \lambda} (\lceil L^1_{ij}/g \rceil, \lceil L^2_{ij}/g \rceil, \lceil \max(\sigma(s),\tau(s))/g \rceil) \quad (1)$$

Because the number of ADM's at a node should be not less than $\max((\sigma(i),\tau(i))/g)$, the minimum number of ADM's should be the total minimum number of ADM's at each node, and is given by

$$M_{\min} = \sum_{i=0}^{n-1} \lceil Max(\sigma(i),\tau(i))/g \rceil \quad (2)$$

*2.3.2 Upper Bounds on irregular networks*

In [14], the author derived the upper bounds on the number of ADM's and wavelengths in tree networks. He assumed that for each node pair a dedicated wavelength was used, so the maximum number of wavelengths was given by $W'_{\max}=n(n-1)/2$.

The upper bound on the number of ADM's was derived by two different ways: the first is to use a dedicated wavelength to support all the traffics. This is equivalent to assigning two ADM's for each traffic demand, then all the traffic demands can be surely originated or terminated at the node even when the traffic patterns vary greatly. It is obvious that for the all-to-all traffics in a network, it requires $n \times (n-1)$ ADM's. The second is to drop the minimum number of wavelengths at every node to accommodate all the traffic demands. The upper bound of ADM's is the smaller one of the two, and given by

$$M_{\max} = \min\{n(n-1), n \times W_{\min}\} \quad (3)$$

But we find that those bounds are not tight enough. It is obvious that using a dedicated wavelength for each node pair in irregular network will lead to consume too many wavelengths. In the following we will try to derive tighter upper bounds on the number of both wavelengths and ADM's.

As for the upper bound on the number of wavelengths, we notice that if each traffic going through the same link is assigned to a different wavelength, no matter how violently the traffic pattern changes, the wavelength can surely accommodate the traffic demands passing through the link. Then the maximum number of wavelengths is equal to the maximum number of calls carried in all the links. Because the number of calls is independent of the number of traffic patterns, we do not need to consider the influence of the traffic patterns.

Assume there are $k$ separate links in an irregular network which we labeled from 0 to $k$-1, let $\omega_i$ denotes the number of traffic demands passing through link $i$, then the maximum number of wavelengths is given by

$$W_{max} = \max_{i=0,\ldots,k-1} \omega_i \tag{4}$$

This upper bound can be applied to all topologies, including ring networks. When discussing a certain topology, it may take on a simpler form. In the following we will derive the bounds on tree and star networks in detail.

Assume there are $n$ nodes in an irregular tree network, and we divide nodes uniformly into the $k$ branches from the root node on the tree. Considering the root node, there are $\lfloor (n-1)/k \rfloor$ nodes on each branch at most, and these nodes will set up lightpaths with $n - \lfloor (n-1)/k \rfloor$ nodes including the root node. So the maximum number of traffic demands passing through each link is $T_{max} = (n - \lfloor (n-1)/k \rfloor) \times \lfloor (n-1)/k \rfloor = \lfloor (n \times (k-1) + 1)/k \rfloor \times \lfloor (n-1)/k \rfloor$. By assigning a wavelength for each of the maximum traffic demand, we can easily get the maximum number of wavelengths required on the tree, which is given by

$$W_{max} = \lfloor (n \times (k-1) + 1)/k \rfloor \times \lfloor (n-1)/k \rfloor \tag{5}$$

For binary tree networks, $k=2$, it is clear that the upper bound on wavelength is

$$W_{max} = \begin{cases} (n^2 - 1)/4 & n \text{ is odd} \\ n^2/4 & n \text{ is even} \end{cases} \tag{6}$$

The bound has a much simpler form for star networks. By regarding the star networks as a simple tree, we can get the upper bound on the number of wavelengths in star networks according to Eq. (5). Assume that there are $n$ nodes in a star network, then there are $n$-1 links connected with the hub (root) node, so $k=n$-1. Then the bounds can be derived as

$$W_{max} = n\text{-}1 \tag{7}$$

We can also derive the bound by analyzing the property of the star networks: the traffics will pass two physics links at most in star networks, and each link will carry $n$-1 traffic demands. By assigning a wavelength for each call, the maximum of wavelengths is equal to $n$-1, which is the same as Eq. (7).

It is obvious from the derivations above that, when the number of nodes in the network is the same, the more the links are connected to the root node (it becomes the central node in star network), the fewer the number of wavelengths is required. For star networks, the number of wavelengths reaches its minimum. Furthermore, in star networks, since all the traffic demands can be assigned to only a few wavelengths, a tighter bound on the number of ADM's will be $M_{max} = n \times W_{min}$. For tree networks, on the contrary, the tighter bound may as well be given by $M_{max} = n(n\text{-}1)$.

As the Equations (1-4) are derived for general cases, they can be applied to all topologies with or without traffic split, including the ring networks. Combined to Equations (5-7), which are derived for the tree and star networks, we can evaluate the computer simulation results objectively.

## 3. Genetic Algorithm Approach

In this paper, we will propose a GA approach to tackle the strictly nonblocking grooming of dynamically changing traffic in tree and star networks. For these two networks, the GA framework is the same but decoding approaches are different. This algorithm structure is flexible and can be easily adapted to the grooming of traffic with traffic split by adjusting the judging conditions when packing traffic into a wavelength.

The framework of this GA approach is based on that described in [5]. We use the ($\mu+\lambda$)-strategy to produce offspring. The outline of this algorithm is described below:
1. Set all necessary parameters, and set $t=0$;
2. Generate an initial population $P_0$ with $\mu$ different chromosomes at random;
3. REPEAT:
    a. Apply crossover and mutation to the parents to produce $\lambda$ offspring;
    b. Decode each chromosome $i$ ($i=1, 2, \ldots, \mu+\lambda$) with Algorithm I to assign each traffic demand in the $M$ traffic patterns to wavelengths in a strictly nonblocking way;
    c. Evaluate each individual $i$ ($i=1, 2, \ldots, \mu+\lambda$) in both the parents and the offspring, and select $\mu$ individuals with the highest fitness value for the next generation from the ($\mu+\lambda$) parents and offspring;
    d. Set $t=t+1$;
4. Until some "termination criterion" is satisfied.

The implementation of each step is described in details in the following subsections.

*3.1 Chromosome Representation*

We first convert the $M$ traffic matrices into $M$ $n(n-1)$-dimensional vectors, $X^m = (x_1^m, \ldots, x_k^m, \ldots, x_{n(n-1)}^m)$, $m=1, 2, \ldots, M$. Then we generate a random permutation of $N=n(n-1)$ different integers in the range [1, $N$] to represent a random permutation of the traffic elements. This permutation will be decoded with the $M$ traffic matrices in a strictly nonblocking manner described in Algorithm I.

*3.2 Decoding approaches and fitness assignment*

In strictly non-blocking grooming, the same call in different traffic patterns must be assigned to the same wavelength when traffic changes. So we decode the chromosome with each of the $M$ traffic patterns one by one. It is obvious that if a wavelength drops at $\alpha$ nodes, it can accommodate at most $\alpha(\alpha-1)$ traffic demands. By packing all these traffics into the same wavelength, we can get the minimum number of ADM's and wavelengths. So we will propose a first-fit approach incorporated with a greedy improvement to decode chromosomes.

In this approach, a traffic item $x_k$ is assigned to a wavelength only when all the traffic demands represented by it in the set of $M$ traffic patterns can be assigned to it. In [5], the first encountered traffic $x_k$ in the chromosome was examined during decoding. If a traffic can not be assigned to the current wavelength, an improvement algorithm is called to examine the remaining traffic items one by one until a new traffic $x_l$ is found which can be assigned to it without adding additional node, that is, both the source and the termination nodes of the traffic $x_l$ have already been equipped with ADM's for that wavelength. If such a traffic item $x_l$ is found, it is assigned to the current wavelength, and the two items ($x_k$ and $x_l$) are exchanged and then the algorithm proceeds with the next wavelength. But if all the items have been examined without finding such a traffic item, the algorithm just stops accommodating the current wavelength and then begins with the next wavelength. This process proceeds until all the traffic items in the chromosome have been assigned to appropriate wavelengths.

The algorithm can be applied to optimizing the traffic assignment in irregular networks after the way of traffic assignment is adapted according to the network structure. We find that when decoding a chromosome, it is necessary to reuse the existing wavelengths in tree networks. That is, before $M$ traffic demands represented by traffic item $x_k$ are assigned to the current wavelength, we need to check whether the existing wavelengths can accommodate them or not because there may be some spare links even after the improvement algorithm proposed in [5] has searched all the traffic items in the chromosome and assigned the right ones to the existing wavelengths. The same can be applied to the star or mesh networks. As wavelength reuse is necessary in irregular topologies, we combine the wavelength reuse technique with the improvement algorithm, and the new algorithm would utilize the capacity of a wavelength much more efficiently. Based on this analysis, we propose an algorithm to decode the chromosome with wavelength reuse technique.

After taking wavelength reuse into consideration, it is impossible to assign traffics to the existing wavelengths without adding additional ADM's. It is obvious that if we always allow to add two additional ADM's when necessary, it would lead to consume too many ADM's. So we try to insert a traffic demand to the existing wavelengths with an additional ADM. Comparing the results with or without wavelength reuse, we found that the algorithm with wavelength reuse would get worse results than the original algorithm occasionally because the search space of the former would be much larger than the latter in some special case. But our simulation showed that by reusing wavelength, we can further save wavelengths and ADM's usually and the average results are always better than without it in each situation. In most cases, the former results are much closer to the optimum ones. For example, when there are 2 traffic patterns and $g$ is 24 in a 15 nodes tree network, with and without wavelength reuse, 115 ADM's and 119 ADM's, 26 wavelengths and 27 wavelengths are required respectively. So reusing wavelength can save 4 ADM's and 1 wavelength. Based on this observation, we always apply this technique when grooming traffic in star or tree networks. Algorithm I implements this algorithm.

*Algorithm I* (decoding a chromosome)

Step 1. $k=0$ and $w=1$;
Step 2. If each of the $M$ traffic demands represented by traffic item $x_k$ can be assigned to wavelength $w$, assign each of them to it. Otherwise, go to Step 9;
Step 3. $f=1$, $l=k+1$;
Step 4. If each of the $M$ traffic demands represented by traffic item $x_l$ can be assigned to wavelength $f$ with an additional ADM added to it, assign each of them to it, $l=l+1$, exchange $x_k$ with $x_l$, $k=k+1$;
Step 5. $f=f+1$, If $f<w$, go to step 4;
Step 6. If each of the $M$ traffic demands represented by traffic item $x_l$ can be assigned to wavelength $w$ without additional ADM added to it, assign each of them to it, $l=l+1$, exchange $x_k$ with $x_l$, $k=k+1$;

Step 7.  $l=l+1$. If $l< n*(n-1)$, go to Step 6;
Step 8.  $k=k+1$. If $k< n*(n-1)$, go to Step 2;
Step 9.  $w=w+1$, go to Step 2;

In this Algorithm, Steps 3 - 5 reuse wavelength by searching each existing wavelength, and other steps perform the search function as mentioned above.

The results of Algorithm I can be served as a solution for strictly nonblocking grooming problem in star and tree networks. When applying this algorithm to the star and tree networks respectively, we must take different steps to assign a traffic to a wavelength which is described in the former section.

After assigning the $M$ traffic patterns to wavelengths, an individual's fitness value is determined by the numbers of ADM's and wavelengths. The one that requires fewer ADM's gets higher fitness value. If two individuals require the same number of ADM's, the fitness value will be decided by the number of wavelengths, and the one requiring fewer wavelengths is assigned higher fitness value. If the number of wavelengths is the same again, a same fitness value will be assigned to them. For the individuals in the next generation, only the µ individuals with the highest fitness value will be selected.

Since we always keep the individuals with the highest fitness value for the next generation, as the iteration goes on, the local improvement used in this approach will finally lead the whole population evolving toward the global optimum rapidly. The effectiveness of this algorithm is demonstrated by the computer simulation results given in the next Section.

*3.3 Crossover and Mutation*

After crossover or mutation, the offspring will be different from the parent chromosome. Then after decoding it with Algorithm I, a new chromosome will be produced. In the process of crossover, we used the operator proposed by Xu and Xu [15], i.e., the Order-Mapped Crossover (OMX) operator. This operator guarantees that no illegal offspring will be produced and the offspring is able to preserve the ordering message from their parents.

In mutation, we adopt the simple inversion mutation, just randomly selecting two points in a parent, then producing an offspring by inversing the genes between the two points.

## 4. Computer Simulations and Discussions

The performance of the proposed strictly nonblocking grooming algorithm was tested in star and tree networks with different granularity and different number of traffic patterns. The traffic demands from node $i$ to node $j$ ($i, j =0, 1, …, n$-1 ) are the random integers distributed uniformly in the interval [0, 15]. The numbers of traffic patterns are 1, 2, 4, and 8 respectively. In all the simulations, the population and the offspring sizes are both set to 200 and the algorithm stops after 500 generations. We tested 10 times for each problem with different randomly generated initial populations, and only the best results are shown. The crossover and mutation rates are set to 0.6 and 0.4 for every individual, and operate in consequence. Two random traffic patterns $\mathbf{R}^1$ and $\mathbf{R}^M$ are generated to represent two extreme traffics, then all the other traffic patterns $\mathbf{R}^m$ (1<$m$<$M$) are generated with $r_{ij}^m$ =random[$r_{ij}^1$, $r_{ij}^M$], which represents the traffic demand originating from node $i$ and terminating at node $j$. For both the star and tree networks, when the number of nodes are the same, the two extreme traffic patterns $\mathbf{R}^1$ and $\mathbf{R}^M$ are the same in all the tests for comparison purpose.

If the varying traffic patterns are treated as static, we must groom the max-traffic matrix to support all the traffic patterns, and the results are always worse than those of our algorithm, especially when the number of traffic patterns is large. The corresponding results are compared in the following discussion.

Fig. 1 shows the grooming results versus the number of nodes on trees with different traffic patterns. Figs. 1(a) and 1(b) give the number of ADM's and wavelengths versus the number of nodes with the traffic granularity $g$=16; Figs. 1(c) and 1(d) give the same results with $g$=24. As the upper bounds will not change when the traffic patterns vary, they are shown in these four figures. However, the lower bounds will change according to the number of traffic pattern, we only show the lower bounds when $g$=24 and there are 8 traffic patterns on the tree respectively in Figs. 1(e) and 1(f). The results of the max-traffic matrix are also shown at the same time to illustrate the advantage of our algorithm.

We can see from these figures that with the increase of the number of traffic patterns, the numbers of both ADM's and wavelengths required on the tree generally increase and gradually incline to the results achieved from the upper bounds. When $g$=16 and there are 8 traffic patterns in the tree, the number of ADM's is very close to the upper bounds. It is because when $g$ is small there are very few ADM's per wavelength in average even after adding proper ADM's to it. Under this condition, the probability of assigning all the traffic demands between a node pair at different moments to a wavelength will be very small, especially when the traffic varies greatly from time to time, such as in the case of 8 traffic patterns. So it is hard to get good savings with strictly nonblocking grooming. But when $g$ is equal to 24, the algorithm performs much better as shown in Figs. 1(c) and 1(d). Furthermore, from Figs. 1(e) and 1(f), we can clearly find that even with as many as 8 traffic patterns, either the results of grooming of the max-traffic matrix or those of our algorithm are smaller than the upper bounds. It is obvious that although grooming the max-traffic request can lead to fewer ADM's and wavelengths in tree networks, the grooming of multiple traffic patterns is always much more beneficial if the $g$ is not too small, especially when the traffic changes quite frequently. What is more, the results of our algorithm are not far away from the lower bounds especially when there are not too many nodes in the tree at the same time. So in the case of traffic varying greatly, when $g$ is not too small, better results can also be achieved. On the other hand, our algorithm is applicable for large scale networks.

Fig. 2 gives the computer simulation results for different granularity when there are 2 traffic patterns and 15 nodes in trees. It shows that with the increase of $g$, the number of ADM's and wavelengths decreases. It can be found that when the $g$ is large, the difference between upper and lower bounds becomes greater without bifurcation. For example, when g=96, although there are only 5 wavelengths in the network all together, each wavelength can carry a large number of traffic requirements, and there are enough spare link capacity to be shared. As a result, more ADM's can be saved when $g$ is large. At the same time, the number of required wavelengths will be very close to or even reach the lower bounds.

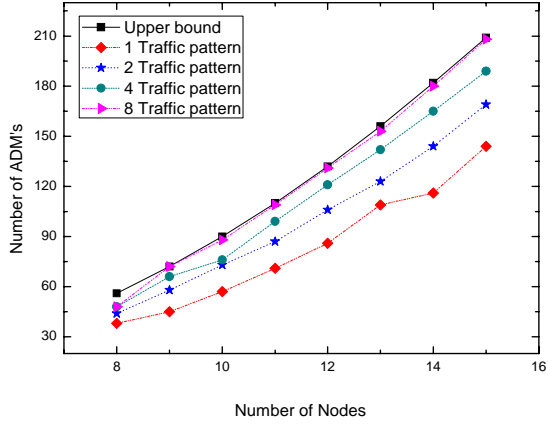

(a)

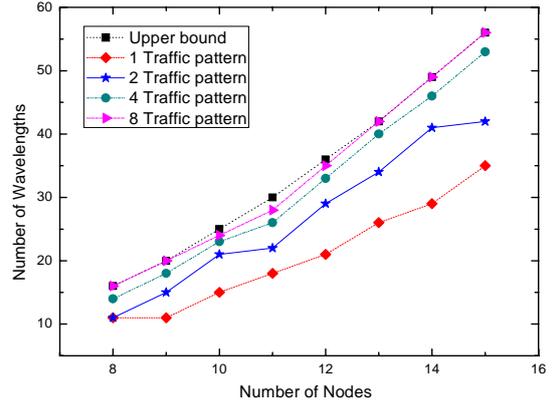

(b)

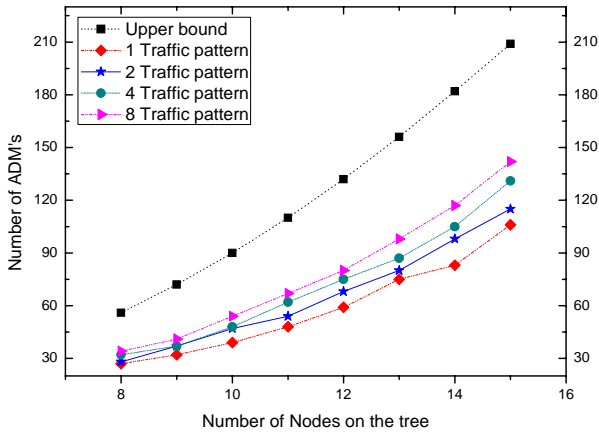

(c)

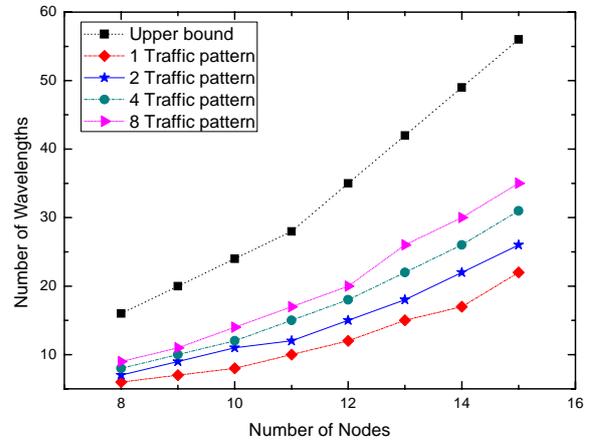

(d)

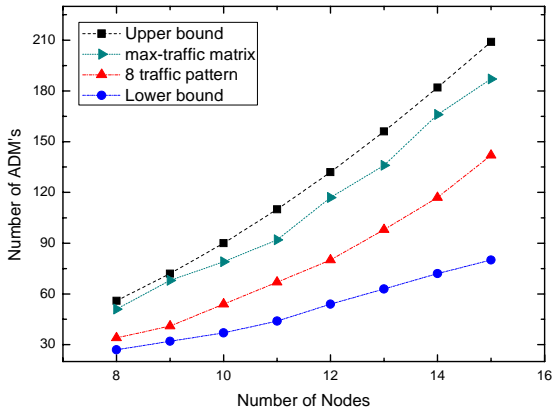

(d)

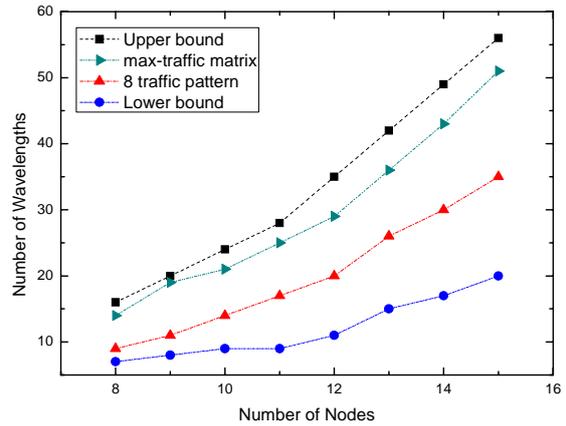

(f)

Fig 1. Computer simulation results for different number of nodes when there are 1, 2, 4, 8 traffic patterns in trees respectively. (a) The number of ADM's vs. nodes for g=16 along with upper bounds; (b) The number of wavelengths vs. nodes for g=16 along with upper bounds; (c) The number of ADM's vs. nodes for g=24 along with upper bounds; (d) The number of wavelengths vs. nodes for g=24 along with upper bounds; (e) The number of ADM's vs. nodes for g=24 along with lower bounds and maximum traffic matrices; (f) The number of wavelengths vs. nodes for g=24 along with lower bounds and maximum traffic matrices;

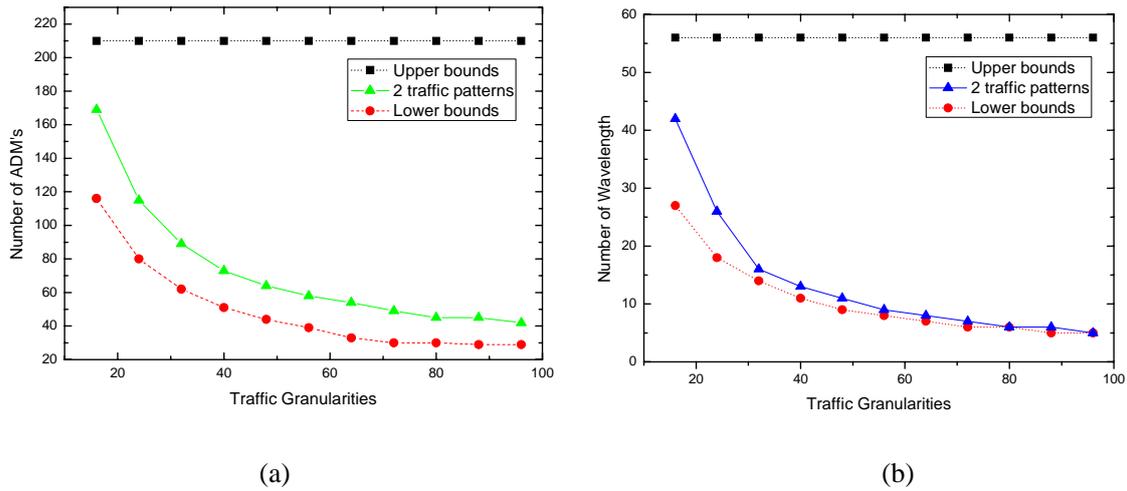

(a)                        (b)

Fig 2. Computer simulation results for different granularity when there are 2 traffic patterns and 15 nodes in trees. (a) The number of ADM's vs. nodes; (b) The number of wavelengths vs. nodes;

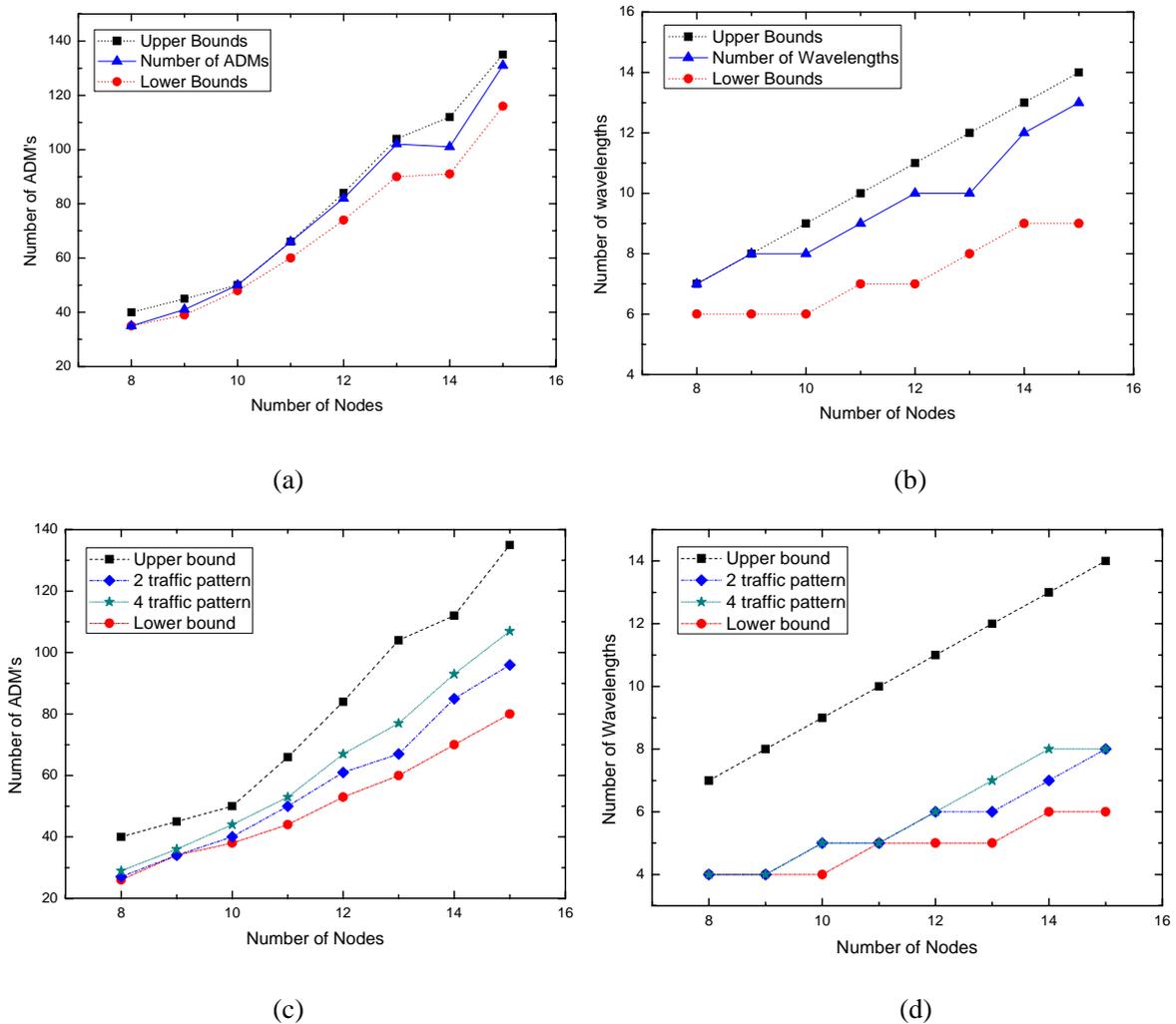

(c)                        (d)

Fig 3. Computer simulation results for different number of nodes in stars. (a) The number of ADM's vs. nodes for g=16 when there are 2 traffic patterns; (b) The number of wavelengths vs. nodes for g=16 when there are 2 traffic patterns; (c) The number of ADM's vs. nodes for g=24 when there are 2 and 4 traffic patterns respectively; (d) The number of wavelengths vs. nodes for g=24 when there are 2 and 4 traffic patterns respectively;

Fig 3. gives the simulation results for different number of nodes with 2 and 4 traffic patterns in star networks. Similar conclusions can be drawn from Fig. 3 for star networks. As shown in Figs. 3(a) and 3(b), when the granularity is 16, the number of ADM's and wavelengths is close or equal to the upper bounds. This is because of the violent changes of the traffic. For example, when the number of nodes is 15 and there are 2 traffic patterns, there are 108.9 traffic demands in average in each link for the first traffic pattern, and 98.9 traffic demands for the second. What is more, the two directions of the fibers in the bidirectional tree carry quite different traffic request. For example, for the second traffic pattern the fiber in direction 0 carries 100.4 traffic demands, while the fiber in direction 1 carries 97.5. Such unbalance would inevitably lead to consume more ADM's. Furthermore, we can see from this figure that the upper bounds and lower bounds on ADM's are very close to each other. As a result, it is difficult to save more ADM's. So when there are 4 or more traffic patterns, all the results reach the upper bounds and we can not save any ADM's or wavelengths. Without bifurcation, we can hardly achieve any better results in saving more ADM's and wavelengths with strictly traffic grooming when the traffics vary greatly.

But when $g$ is equal to 24, more ADM's and wavelengths can be saved. As the lower bounds on both ADM's and wavelengths for 2 and 4 traffic patterns are equal in almost each case, we show the lower bounds for 4 traffic patterns to evaluate the results. It is clear that the number of ADM's and wavelengths are all very close or equal to the lower bounds when there are 2 traffic patterns. Our simulation showed that when there are 15 nodes in a star network, the average link load is 13.6 for the first traffic pattern, 12.3 for the second with only 10 ADM's in each wavelength on average. We think that our results are quite satisfactory for the strictly nonblocking grooming. And with the increasing of number of nodes, the number of saved ADM's and wavelength is much larger. Even when there are 4 traffic patterns and 15 nodes in the star, we can still save 28 ADM's and 6 wavelengths at most. So it proves that our algorithm can be applied to large scale star networks.

The results in star networks may not be as good as those in tree networks according to Figs. 1 to 3. But it is reasonable since in a star network with fully packed wavelengths, there are not enough spare link load to utilize. What is more, each wavelength drops at 10-11 nodes in average in all the case when there are 15 nodes on the network, so the saving on ADM's is more difficult than in tree networks, but the results are still much better than grooming of max-traffic matrix which never achieves any saving.

Fig. 4 depicts the simulation results with different granularities in a star network with 15 nodes and 2 traffic patterns. We can see from this figure that when $g$ is large enough, the algorithm can achieve good saving. Specifically, with the increase of the granularity in star networks, the number of ADM's and wavelengths decrease and become close to the lower bounds. When the granularity is far larger than the maximum traffic request, for example, g>=72, which is 4 times larger than the maximum traffic request in this network model, the upper bound on the number of ADM's become equal to the lower bound, which means that no grooming algorithms can further save ADM's in this case. We can see from Fig. 4(a) that this happens when g=48. The reason is that in the star networks, for each leaf node $i$, $\max(\sigma(i), \tau(i)) = \sum_{j=0,...,n-1; j \neq i} (\max(L^1_{ij}, L^2_{ij}))$. By labeling the hub node as 0, the minimum number of wavelengths in the star networks is given by

$$W'_{\min} = \max_{i=1,...,n-1} (\lceil \max(\sigma(i), \tau(i))/g \rceil, \lceil \max(\sigma(0), \tau(0))/g \rceil) \qquad (8)$$

According to the discussion above, in star networks, the maximum and minimum number of ADM's is $M_{\max} = n \times W'_{\min} = n \times \max_{i=1,...,n-1} (\lceil \max(\sigma(i), \tau(i))/g \rceil, \lceil \max(\sigma(0), \tau(0))/g \rceil)$, and $M_{\min} = \sum_{i=0}^{n-1} \lceil Max(\sigma(i), \tau(i))/g \rceil$. So with the increasing of $g$, $\lceil \max(\sigma(i), \tau(i))/g \rceil$ and $\lceil \max(\sigma(0), \tau(0))/g \rceil$ becomes more and more closer, and the difference between the upper bound and the lower bound of ADM's will be smaller and smaller. Finally, when $g$ is large enough, the difference between them vanishes, which comes $M_{\max} = M_{\min}$.

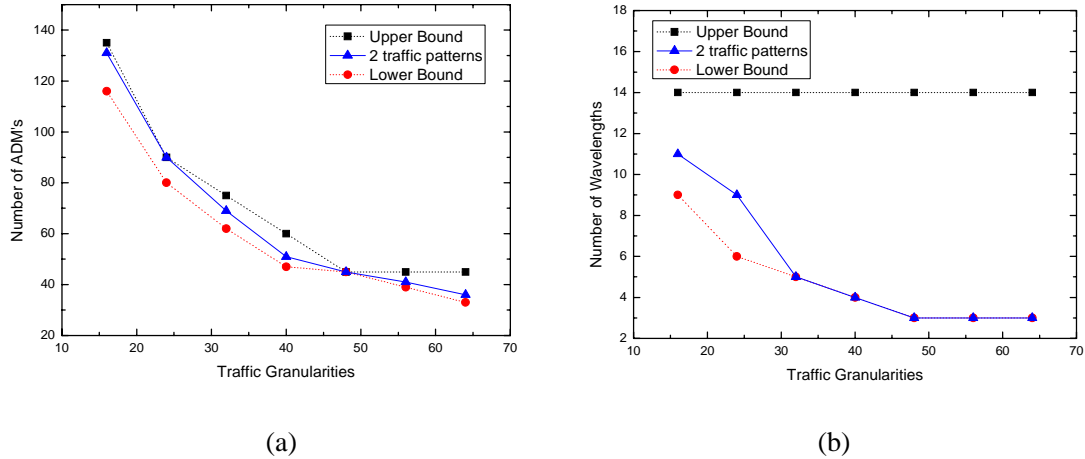

(a)            (b)

Fig 4. Computer simulation results for different granularity when there are 2 traffic patterns and 15 nodes in stars. (a) The number of ADM's vs. granularities; (b) The number of wavelengths vs. granularities;

Although it is generally impossible to reach the lower bounds for arbitrary traffic patterns as analyzed in [5], the results are not far from the lower bounds especially when both the numbers of nodes and traffic patterns are small and $g$ is large enough. With the increasing of either the number of traffic patterns or the number of nodes, it is harder and harder for the results to reach their lower bounds.

From all the given figures, we can find that our algorithm performed steadily with the change of granularity and nodes in tree and star networks.

## 5. Conclusion

We have introduced a GA approach to tackle the problem of strictly nonblocking grooming of dynamic traffic on the tree and star networks in this paper. To evaluate the performance of the algorithm, we derived tighter bounds on ADM's and wavelengths. Since strictly nonblocking grooming will assign the same traffic demands in different traffic patterns to the same wavelengths, it may assure that the setup of lightpath would not be interrupted all the time, and will result in a more efficient use of wavelengths' capacity than just grooming the maximum traffic. Computer simulation results showed that our algorithm can achieve very good results in saving both ADM's and wavelength in star or tree networks. In addition, our algorithm can be applied to any kind of tree networks.

Because rearrangeably nonblocking grooming may lead to save more ADM's and wavelengths, discussion on it would be our future work. What is more, with bifurcation, better results could be achieved [6, 16], and this is also an important aspect we should try.